# Do we really need to recalibrate the CT system Configuration for every experiment?

Kajal Kumari[1$], Mayank Goswami[1#],

[1]Divyadrishti Imaging Laboratory, Department of Physics, IIT Roorkee, Roorkee, India
Department of Physics, IIT Roorkee, Roorkee, India

[$]kkumari@ph.iitr.ac.in

[#]mayank.goswami@ph.iitr.ac.in,

## Abstract

Different technical and physical factors may affect the image quality reconstructed using a Computed Tomography system. We have developed and designed a 2-D Gamma Computed Tomography set up to study the effect of some physical parameters. One must decide the number of detectors and set CT geometry parameters/ configuration like the fan-beam angle and number of rotations. Usually, geometry parameters are determined based on the object's size. This study shows the influence of the density distribution of same-sized phantom/objects on CT geometry parameters.

Due to limited space, industrial applications may not allow a CT system to move around the object (under investigation). On-spot customization of CT system configuration may be required according to similar situations. The same problem is experienced in medical science, material science, and many other fields in which the CT system is widely used for non-destructive imaging. The number of detectors in the scanning array is one major factor that optimizes a CT system. Of course, more detectors are desired for better resolution. Changing the number of detectors requires recalibration of CT geometry.

A simulated work is presented to study the influence of the density distribution of same-sized objects and the number of detectors on CT system configuration. The same is verified experimentally also. A comparison between simulated and experimental results shows a good agreement.

## Introduction

Computed Tomography has been extensively used as non-invasive imaging in medical diagnosis and surgical planning and as non-destructive testing in industrial applications. In practical applications of computed tomography, sometimes limited data is obtained due to restrictions in data collection time or/and constraints on CT Geometry[1]. The data collection time restricts the user from selecting the number of rotations. The constraints on CT geometry limit the source to detector distance and so on the fan-beam angle and the number of detectors. The change in geometry parameters requires on-spot optimization of CT geometry. Multiple experiments cannot be performed to find out the optimal CT geometry. It raises the patient's risk of exposure to radiation in the medical field[2]. On the other hand, in the industrial field, time, memory consumption, and changing the distribution of the object with time are the main factors that disapprove the on-spot CT geometry optimization[3].

One important parameter affecting the resolution of the CT images is the number of rotations and detectors. An increase in the number of rotations will improve the CT image quality. However, increasing the number of rotations will also increase the scanning time. For good spatial resolution, the required number of detectors depends on the fan-beam angle, which is decided by the diameter of the scanned object[4].

Usually, people decide the CT geometry parameters based on the object's size and detector's size. Similar work is done in which CT geometry parameters are calculated based on the diameter of the object and detector and length of the detector[5]. Some people choose the CT geometry parameters according to their own without any clear explanation[6], [7].

In our work, we have selected two heterogeneous phantoms of the same size. A simulation study is done to evaluate the influence of the density distribution of same-sized objects and the number of detectors on CT geometry. The results obtained by simulations are validated with experimental results.

## Motivation

The motivation of this work is to show the dependency of the density distribution of same-sized objects on CT geometry parameters. This study shows why it is essential to recalibrate the CT geometry for every single object. In this work, we also evaluated that changing the number of detectors can degrade the CT image quality for an optimized CT geometry. This change requires recalibration of CT geometry.

## Materials and Methods

### Theory

This study shows a general dependency of density distribution and the number of detectors on CT geometry parameters, which can be visualized using ionized or/ and non-ionizing radiation-based computed tomography. The simulated results are verified with experimental results using a gamma CT scanner.

### Simulation Study

The simulation study is carried out using the filtered back projection reconstruction algorithm. Our MATLAB codes are written to evaluate the simulated optimal CT geometry parameters for every phantom. The geometry settings that give minimum root mean square error (RMSE) of simulated reconstructed image w.r.t. phantom is considered optimal CT geometry.

### Experimental Details

The experiment comprises a gamma radioactive source Cs-137 of activity 1.5μCi, heterogeneous phantoms, and a gamma-ray detector, as shown in fig.1. A NaI (Tl) scintillator crystal coupled with its photomultiplier (PM) tube (make: electronics enterprise Ltd. India) is considered a gamma ray detector. Its anode output is amplified by the separate amplifier circuit, which is controlled via a single-channel analyzer (SCA). Gamma CT experiment is performed at optimal CT geometry obtained by simulations.

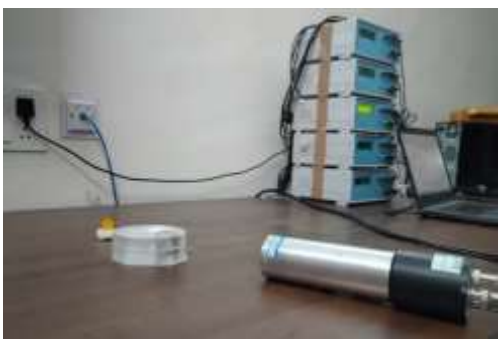

Fig.1: Experimental setup of gamma CT scanner.

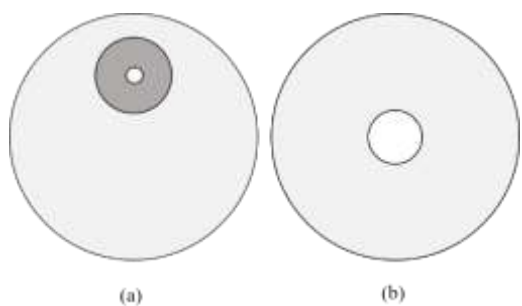

Fig. 2: Schematic diagram of (a) phantom1 and (b) phantom 2.

For scanning purposes, two well-known cylindrical-shaped phantoms are used. These phantoms are made up of Perspex material with a diameter of 12 cm. In phantom 1, two holes of diameter 3.8 and 0.8 cm are drilled off-centered and afterward filled with concentric aluminum and iron cylinders, respectively. In Phantom 2, a hole diameter of 2.6 cm is drilled centered and afterwards filled with an iron cylinder, as shown in fig.2.

## Results and Discussion

### Influence of Density Distribution on CT Geometry

Ideal projection data is obtained and then reconstructed using a filtered back projection reconstruction algorithm. The simulated results do not contain any electronic noise, background noise, and non-linear error due to scattering, it includes only reconstruction error. The optimal CT geometry parameter: fan-beam angle and number of rotations are predicted using a simulation study. For every phantom, the optimal CT geometry parameters are obtained.

The simulated reconstructed results are obtained using these settings. For a fixed number of detectors, which is used nine, the optimal fan-beam angle and rotations are obtained at 56° and 16 for phantom1 and 40° and 12 for phantom 2. For every phantom, simulated reconstructed results are obtained corresponding to the optimal setting of phantom1 and phantom2 as shown in fig. 3. RMSE values 9.414%, 11.445%, 11.923%, and 17.088% are obtained corresponding to simulated reconstruction results shown in fig. 3 (b), (c), (d) and (f) respectively.

Gamma CT experiment is performed with these settings. The reconstructed results are shown in fig. 4.

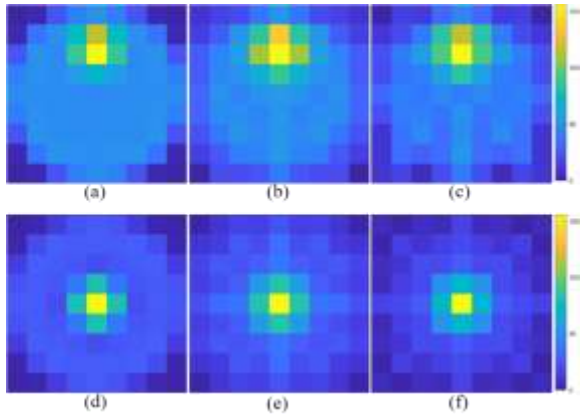
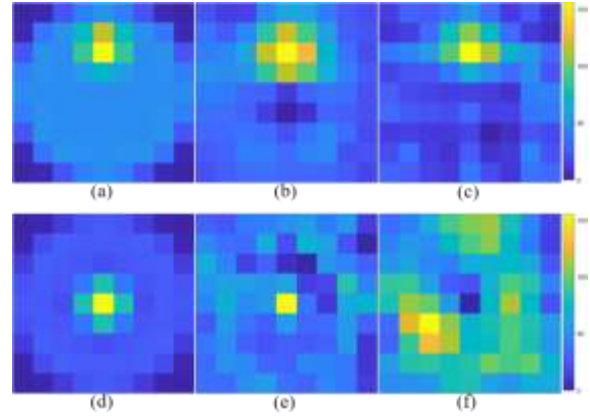

Fig. 3: (a) phantom 1, simulated reconstruction of phantom 1 at optimal setting of: (b) phantom1, (c) phantom 2, (d) phantom2, simulated reconstruction of phantom 2 at optimal setting of: (e) phantom 2 (d) phantom 1.

Fig.4: (a) phantom1, experimental reconstruction of phantom 1 at optimal setting of: (b) phantom1, (c) phantom2, (d) phantom2, experimental reconstruction of phantom2 at optimal setting of: (e) phantom2, (d) phantom 1.

Fig. 4(b) and (c) show the experimental reconstructed images for phantom1 obtained at an optimal setting of phantom1 and phantom2. Fig. 4(d) and (e) show the experimental reconstructed images for phantom2 obtained at an optimal setting of phantom2 and phantom1. RMSE 19.082%, 22.607%, 40.687% and 90.234% is calculated corresponding to experimental reconstruction results shown in fig. 4 (b), (c), (e) and (f) respectively. The experimental results are in good agreement with simulated results. However, a huge difference can be observed in experimental reconstruction results obtained for phantom2.

The results show that for every single object, it is necessary to recalibrate the CT geometry to obtain better qualitative and quantitative information from CT images. The optimal fan-beam angle and number of rotations are different for different density distributed heterogeneous objects. For phantom1, more rotations are required as compared to phantom2. This study verifies that resolving the different density materials present in scanned objects requires more rotations. However, more number rotations do not always need to give better spatial resolution. There should be an optimal value that can be computed by a simulation study. In phantom1, the heterogeneity is present off-centered, while in phantom2, the heterogeneity is present at the center. The position of heterogeneity also affects the number of rotations and fan-beam angle.

**Influence of Number of detectors on CT Geometry**

The influence of the number of detectors on CT images for optimized CT geometry has been investigated. The number of detectors is changed from nine to eight, and then optimal CT geometry parameters are calculated. It is observed that the CT geometry parameters changed with the changing number of detectors. To view the effect of detectors on CT images, we have simulated reconstructed CT images before and after CT geometry, as shown in fig. 5. Gamma CT experiment is performed for phantom1 at optimal CT geometry obtained for nine and eight detectors. The reconstructed results (using experimental data) are shown in fig. 6. RMSE values 9.954%, and 8.554% are obtained for reconstructed results shown in figures 6 (b) and 6(c), respectively. For eight detectors, fan-beam angles kept the same while the number of rotations are changed from 16 to 17. The effect of the rise of only one rotation can be clearly visualized in fig. 6(c). The previous CT geometry settings, fig 6(b) unable to resolve the iron cylinder from Al cylinder and looses the quantitative and qualitative information. These results show that changes in the number of detectors can degrade the CT image quality for an optimized CT geometry.

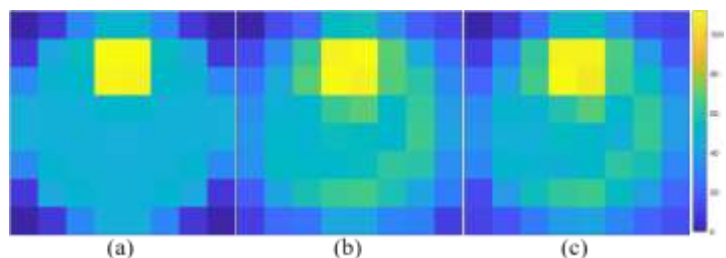

Fig.5: (a) Phantom 1, simulated reconstruction at optimal CT geometry obtained for: (b) nine detectors and (c) eight detectors

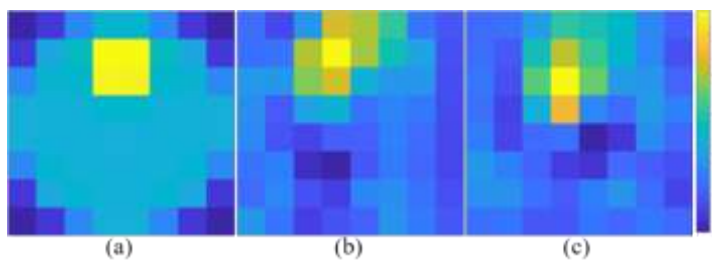

Fig.6: (a) Phantom 1, experimental reconstruction at optimal CT geometry obtained for: (b) nine detectors , (c) eight detectors

## Conclusion

This paper investigates the dependency of the density distribution of same-sized objects and the number of detectors on CT geometry parameters. A comparison of results acquired by simulation and experiment shows that the simulation study helps the user find the optimal CT geometry for real-life experiments. Both the number of rotations and detectors define the resolution of CT images. More rotations increase the scanning time only, but an increase in the number of detectors also increases the cost of the CT system. For a fixed number of detectors, it gives the opportunity for the user to calculate the optimal CT geometry.

The primary outcome of this study is that the number of rotations is dependent on the types of density materials present in the scanned object. More rotations are required to resolve the distinct materials. However, there should be a threshold value, which is calculated by a simulation study. Other simulation tools like Monte Carlo and GEANT4 could be used, but these take the same time as a real experiment. This general study applies to ionizing and non-ionizing radiation-based CT systems.